\documentclass[prl,superscriptaddress,showpacs,reprint]{revtex4-1}

\usepackage{amsmath,amssymb,graphicx,color,bm,soul,ulem}

\begin{document}

\title{Heat-assisted self-localization of exciton polaritons}

\author{I. Yu. Chestnov}
\affiliation{Vladimir State University named after A. G. and N. G. Stoletovs, 87 Gorkii st., 600000 Vladimir, Russia}
\author{T. A. Khudaiberganov}
\affiliation{Vladimir State University named after A. G. and N. G. Stoletovs, 87 Gorkii st., 600000 Vladimir, Russia}
\author{A. P. Alodjants}
\affiliation{ITMO University, St. Petersburg 197101, Russia}
\author{A. V. Kavokin}
\affiliation{CNR-SPIN, Viale del Politecnico 1, I-00133 Rome, Italy}
\affiliation{University of Southampton, Physics and Astronomy School, Highfield, Southampton, SO171BJ, UK}
\affiliation{Russian Quantum centre, 100 Novaya st., 143025 Skolkovo, Moscow Region, Russia}
\affiliation{Spin Optics Laboratory, St. Petersburg State University, St. Petersburg 198504, Russia}


\begin{abstract}
Bosonic condensation of microcavity polaritons is accompanied by their relaxation from the ensemble of excited states into a single quantum state. The excess of energy is transferred to the crystal lattice that eventually involves heating of the structure. Creation of the condensate results in the local increase of the temperature which leads to the red shift of the exciton energy providing the mechanism for polariton self-trapping. By employing the driven-dissipative Gross-Pitaevskii model we predict a new type of a stable localized solution supported by the thermally-induced self-trapping in a one-dimensional microcavity structure. The predicted solution is of a sink-type i.e. it is characterized by the presence of converging density currents. We examine the spontaneous formation of these states from the white noise under spatially localized pumping  and analyze the criteria for their stability. The collective bosonic polaron state described here may be considered as a toy model for studies of bosonic stars formed due to the self-gravity effect.
\end{abstract}

\maketitle

Exciton-polaritons are hybrid quasiparticles arising under the strong coupling of semiconductor excitons and an electromagnetic mode of a microcavity. The  most fascinating  property of polaritons is their spontaneous condensation in  a single quantum state \cite{kasprzak2006,sun2017}. Since the first observation, the condensates of exciton polaritons serve as a powerful tool for exploring  fundamental phenomena where  quantum many-body  physics and  nonequilibrium dynamics meet.

\begin{figure}
\includegraphics[width=1\linewidth]{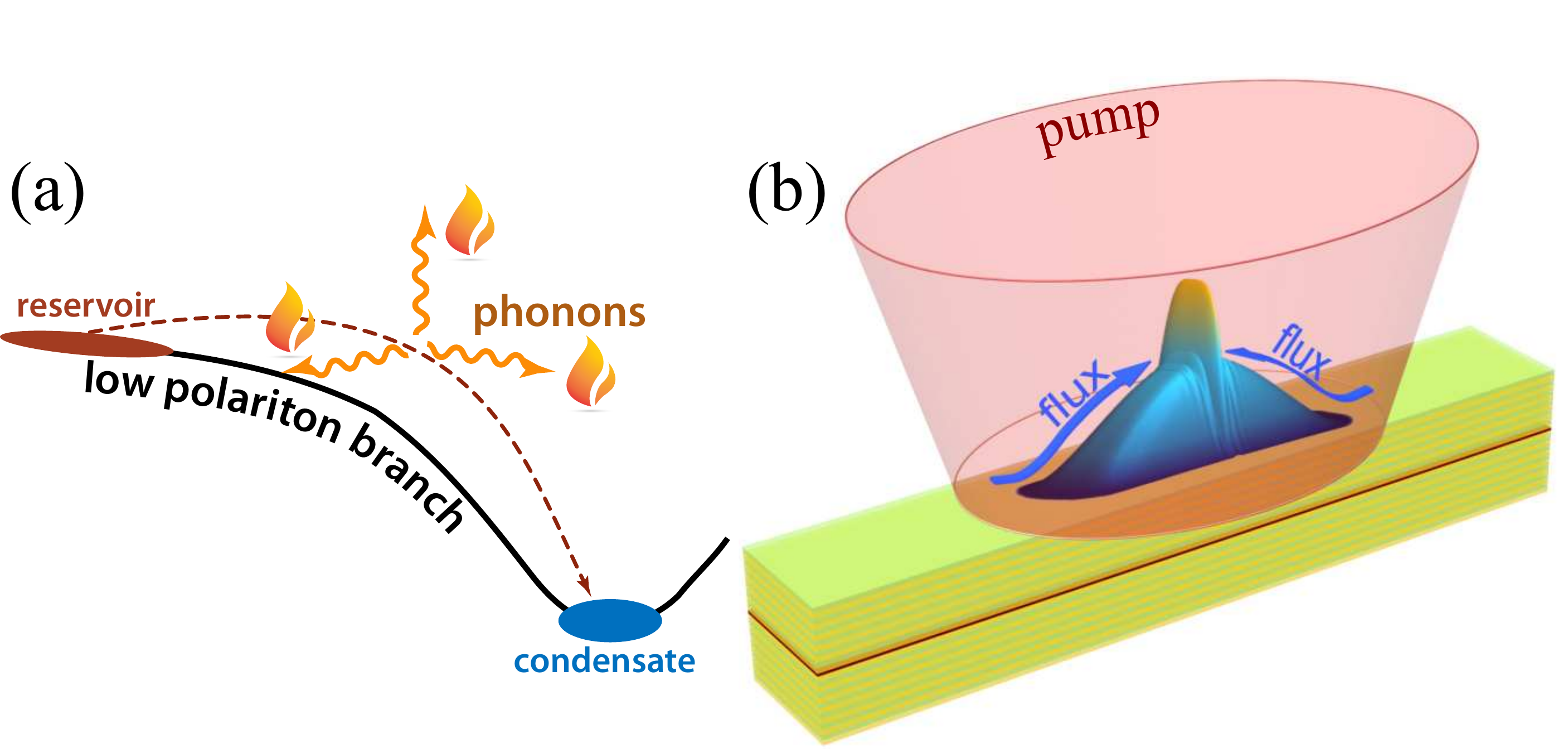}
\caption{(a) The schematic showing the dispersion of the low polariton branch and the phonon-assisted scattering processes leading to heating of the crystal lattice. (b) Sketch of a microcavity stripe excited by a non-resonant pump. The polaron solution is formed as the result of interference of the incoming  polariton fluxes. }
\label{Fig.scheme}
\end{figure}

Because of the strong dissipation polariton condensates can only be formed in the presence of the external pump. We shall specifically address here the experimental configuration that implies a non-resonant optical pumping. The pumping creates a reservoir of non-condensed excitons. Due to the stimulated scattering process these excitons reduce their energy by joining the condensate of exciton polaritons. The excess of energy is dissipated in the crystal lattice, as Fig.~\ref{Fig.scheme}(a) schematically shows.
Heating of the crystal lattice is an unavoidable feature of any experiment with the incoherent excitation of a polariton condensate. Although the threshold of polariton lasing is relatively low, the heating of the sample due to the relaxation of excitons to the exciton-polariton condensate may be rather significant, especially in the continuous wave pumping regime \cite{klembt2015}.

In this Letter, we demonstrate that the heat released during the condensation may help localization of polariton condensates  in the plane of the cavity. We consider the local variation of the crystal lattice temperature due to the emission of acoustic phonons which assist relaxation of hot excitons from the reservoir to the  condensate ground state. The  temperature increase induces the renormalization of the semiconductor band gap resulting in the lowering of the exciton energy and thus favors localization of polaritons. This constitutes a self-trapping mechanism that may lead to a formation of self-localized condensate states that are constantly fed from the reservoir due to the stimulated scattering processes.

The manifestation of the thermally induced condensate self-trapping was experimentally observed \cite{dominici2015} under coherent polariton excitation by short optical pulses. Instead of the anticipated diffusion away from the excitation spot the  collapse of the polariton fluid   into a tight spot has been observed in these experiments. The observed localization was interpreted in terms of the collective polaron effect induced by the local heating of the crystal lattice. The similar self-trapping due to the collective magnetic polaron effect was also predicted for condensates of polaritons in semimagnetic microcavities \cite{shelykh2009,miketki2017} and earlier for excitons in quantum wells \cite{kavokin1993}.

Formation of localized states  of polariton superfluids was demonstrated both theoretically and experimentally in different microcavity systems. The most striking examples are bright \cite{egorov2009,walker2015} and dark \cite{yulin2008} solitons which are formed at the negative and positive effective mass regions of the low polariton dispersion branch, respectively, dissipative solitons and vortices \cite{ostrovskaya2012,ostrovskaya2013} occurring due to the balance of gain and superfluid density flows, etc. In the present work we study theoretically a new type of the localized states: collective bosonic polarons. Formation of a bosonic polaron manifests itself in  the sink-type solution of the Ginzburg-Landau equation.
This new topological state is formed under inhomogeneous nonresonant pump and represents a terminating line connecting counter propagating polariton flows with a bright soliton-like intensity peaks at the point where they meet.

The model system studied in this Letter is schematically shown in Fig.~\ref{Fig.scheme}(b). We consider a one-dimensional condensate of exciton polaritons which is realized, for instance, in the case of trapping of polaritons in a microcavity stripe or a microwire \cite{wertz2012}.
The condensate is excited by a non-resonant continuous wave pump. We apply the mean field approach using the complex Ginzburg-Landau equation for the polariton wave function (the order parameter) $\Psi$ coupled to the rate equation for the density $n$ of incoherent excitons:
\begin{subequations}\label{system}
\begin{eqnarray}
i\hbar \partial_t \Psi &=& \left[ -\left(\frac{\hbar^2}{2m} - i\Gamma_e \right) \partial_{xx}  + g_c|\Psi^2|  + g_r n   \right. \\
\notag &  & + \alpha(\delta T) +\left. \frac{i\hbar}{2} \left( R n - \gamma_c \right) \vphantom{-\frac{\hbar^2}{2m} \Delta} \right] \Psi,\\
 \partial_t n &=& -\left( \gamma_r + R |\Psi^2| \right)n + P(x).
\end{eqnarray}
\end{subequations}
Here $m$ is the polariton effective mass, $g_c$ denotes polariton-polariton repulsion strength while $g_r$ describes the repulsion between condensate polaritons and hot  reservoir excitons. The values of $\gamma_r$ and $\gamma_c$ are the condensate and the reservoir decay rates  which are balanced by the pump $P$ at the steady state. The transfer of reservoir excitons to the condensate state occurs with the rate $R$. The term containing $\Gamma_e$ accounts for the effect of energy relaxation  as described in \cite{bobrovska2014,kulczykowski2015,wertz2012,tanese2013}.

The term $\alpha(\delta T)$ in (\ref{system}a) is responsible for the shift of the low polariton branch induced by the local variation of the lattice temperature $\delta T$. Since the heat originates from the exciton scattering from the reservoir to the condensate, the energy transferred to the crystal lattice is proportional to the rate of scattering events, $R n |\Psi^2|$.
We neglect by the diffusive transport of the heat since it is slow on the picosecond time scale typical for the condensate dynamics. The relaxation of the temperature caused by the heat transfer out of the quantum well region  or due to any other mechanisms  should be balanced by the heating source in the steady-state regime. Thus we assume an instantaneous feedback of the lattice temperature, i.e. we take the temperature variation $\delta T$ in the form $\delta T = \beta R n |\Psi|^2$, where $\beta$ is a phenomenological coefficient. The value of $\beta$ is a key parameter of the polaron formation that is governed by the balance of the heating and relaxation mechanisms. The retardation effect discussed in Ref.~\cite{dominici2015} is neglected here for simplicity.

The value of the polariton energy shift $\alpha(\delta T)$ induced by the temperature variation is dependent on the  properties of a quantum well material. Here we focus on  microcavities based on GaAs for which the energy band gap renormalizes as $\varepsilon_{\rm{g}}(T) = \varepsilon_{\rm{g}}{(T=0)} - \frac{0.541 T^2}{T+204}$ (the energy is measured in  meV) \cite{GaAsBandGap}. We assume that the exciton energy follows the same dependence. The energy of the lower polariton branch  is dependent on the exciton and the microcavity photon energies: $\varepsilon_{\rm{pol}} =  C_p^2 \varepsilon_{\rm{cav}} +  C_x^2 \varepsilon_{\rm{ex}} - 2C_x C_p \Omega $, where $C_{x,p}=\frac{1}{\sqrt{2}}\left(1 \pm \frac{\delta}{\sqrt{\delta^2 + 4\Omega^2}}  \right)^{1/2}$, $\delta= \varepsilon_{\rm{cav}} - \varepsilon_{\rm{ex}}$ is the cavity-exciton detuning and $\Omega$ is the Rabi splitting. Neglecting the dependence of the Hopfield coefficients $C_{x,p}$  on the exciton-photon detuning $\delta$ that is valid at the bottom of the lower polariton branch, we denote $\alpha(\delta T)= \Delta \varepsilon_{\rm{pol}} = \varepsilon_{\rm{pol}}(T)- \varepsilon_{\rm{pol}}{(T=T_0)}  $, where $T_0$ is the lattice temperature in the empty cavity (without condensate). Thus at zero detuning  $\delta=0$, and in the limit of low temperature where both $T_0$ and $\delta T$ do not exceed few tens of Kelvin, one can express
\begin{equation}\label{thermal_nonlinearity}
\alpha(\delta T)\approx -2 \alpha_0 T_0 \delta T  - \alpha_0 \delta T^2,
\end{equation}
where $\alpha_0 = 1.325\times 10^{-3}$ meV$\cdot$ K$^{-2}$. The  sign minus in Eq.~\eqref{thermal_nonlinearity} corresponds to the effective attraction between polaritons which should affect the spectral and the spatial structure of the condensate.

The impact of the heating on the properties of the polariton condensate is investigated below. We are particularly interested in the steady state solution for the polariton density $\Psi=\psi(x) e^{-i\mu t}$ formed in the presence of the spatially inhomogeneous pump having a Gaussian shape:
\begin{equation}\label{pump}
P(x) = P_0 e^{-\left(x/ w\right)^2},
\end{equation}
where $w$ is the pump width.
We simulate the condensate dynamics described by Eqs.~(\ref{system}-\ref{pump}) on the time scale of several nanoseconds (depending on the system parameters) that  is enough for the formation of the steady state. We assume that at $t=0$ the continuous wave pump is switched on and the condensate starts growing from the low amplitude white noise which mimics the thermal fluctuations of the polariton field.
\phantom{\footnote{Parameters: $\gamma_c=0.33$~ps$^{-1}$, $\gamma_r=1.5\gamma_c$, $R=0.0075$~ps$^{-1}\mu$m, $g_c=1.5\times 10^{-3}$meV$\mu$m, $g_r=2g_c$, $m=0.568$ meVps$^2$$\mu$m$^{-2}$. The pump width is $w=120$~$\mu$m.}}

\begin{figure}
\includegraphics[width=0.9\linewidth]{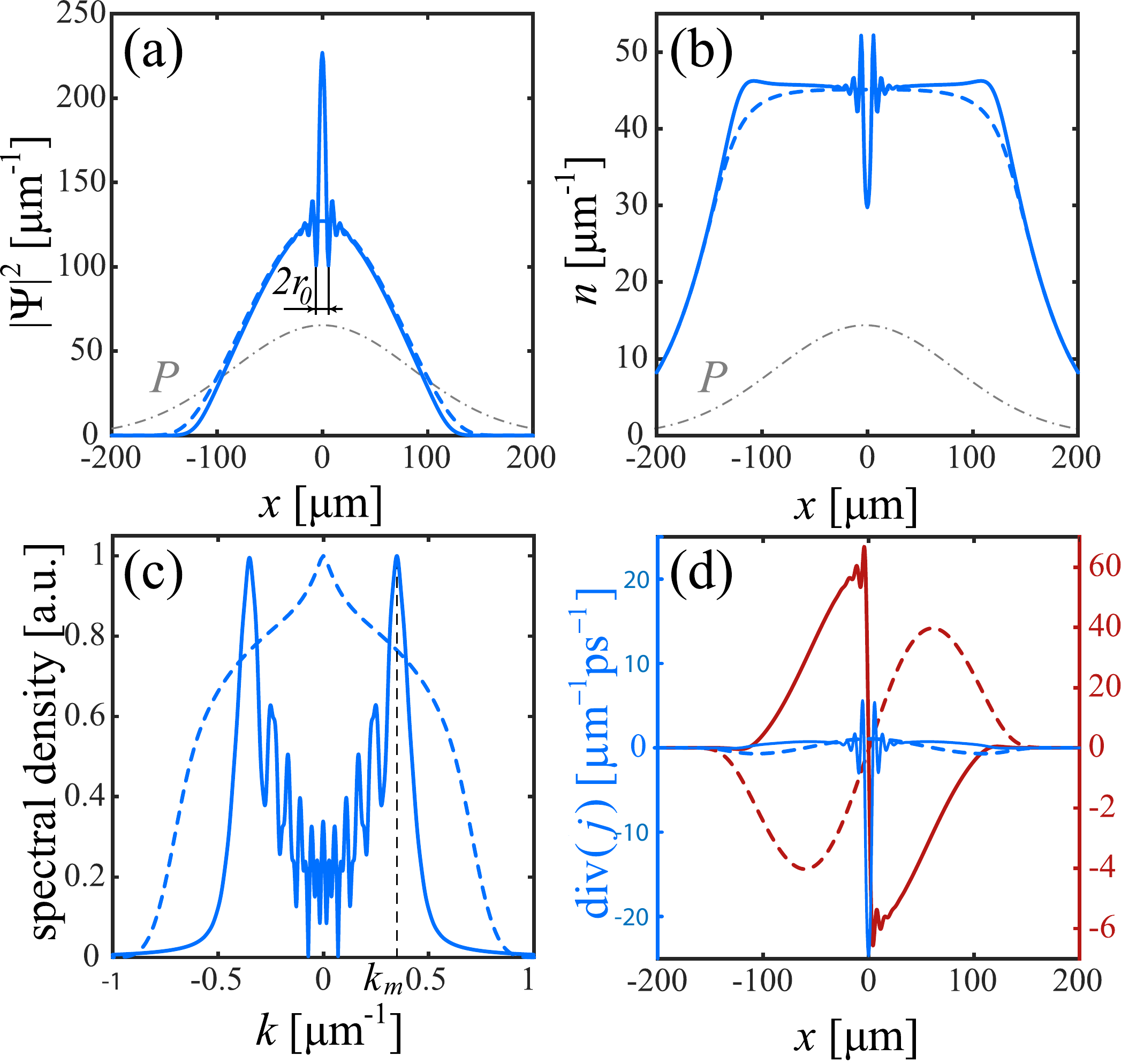}
\caption{The self-localized steady state solution formed by the Gaussian pump, in the absence of thermally induced energy red shift, $\beta = 0$ (dashed curves) and in the presence of it with $\beta = 0.1$ K$\cdot$ps$\cdot \mu$m  (solid curves). (a)  The polariton density $|\Psi|^2$. The dash-dotted curve sketches the pump intensity profile. (b) The distribution of excitons in the reservoir, $n$. (c) The spatial spectra of the solutions shown in the panel (a). (d)  The polariton flux $ j $ (red curves, right axes) and the divergence of the flux ${\rm{div}}(j) = \partial_x j(x)$ (blue curves, left axes). The pump amplitude is $P_0=3 P^{\rm{th}}$, where $P^{\rm{th}}=\gamma_c \gamma_r/R$ is a condensation threshold in the system with a homogeneous pump. Values of the other parameters are given in  \cite{Note1}. }
\label{Fig.formation}
\end{figure}

When heating of the crystal lattice is neglected (dashed curves in Fig.~\ref{Fig.formation}), i.e. $\beta=0$,  the condensate density acquires a ``bell-shaped'' profile \cite{ostrovskaya2012}, see panel (a). In this case, because of the polariton-polariton repulsion the condensate flows outwards  from the pump spot.
The  polariton flux  $j = \frac{\hbar}{m} \Im(\Psi^{\ast} \partial_x \Psi)$  gradually grows with the distance from the centre (panel (d)) and decreases at the periphery of the condensate spot. 

In contrast, once  the thermally induced nonlinear terms \eqref{thermal_nonlinearity} are taken into account, we observe the appearance of a new non-trivial feature in the spatial distribution of the polariton condensate (solid curves in Fig.~\ref{Fig.formation}). Namely, two counter propagating currents flow  towards the centre of the pump spot, see panel (d). These currents interfere at the meeting point forming a stable bright soliton-like pattern with oscillating tails, Fig.~\ref{Fig.formation}(a). This density maximum corresponds to the pronounced collapse  of the flux divergence, ${\rm div} (j) = \partial_x j(x)$,down to negative values, see Fig.~\ref{Fig.formation}(d). Thus the central peak connects two domains of the incoming fluxes and serves as a sink.

The class of sink solutions is described by the theory of complex Ginzburg-Landau equation \cite{saarlos1992}. Besides, the scheme for the observation of a sink-type state of polariton condensate was proposed recently \cite{kulczykowski2015}. The approach of Ref.~\cite{kulczykowski2015}, however, does not account for the effect of heating. Instead, it implies the specifically designed inhomogeneous pump constructed by the long homogeneous region and two high peaks at the edges which generate counter propagating polariton currents. Quite contrarily, in our case this state forms spontaneously  in the presence of a spatially extended Gaussian pump.

\begin{figure}
\includegraphics[width=0.93\linewidth]{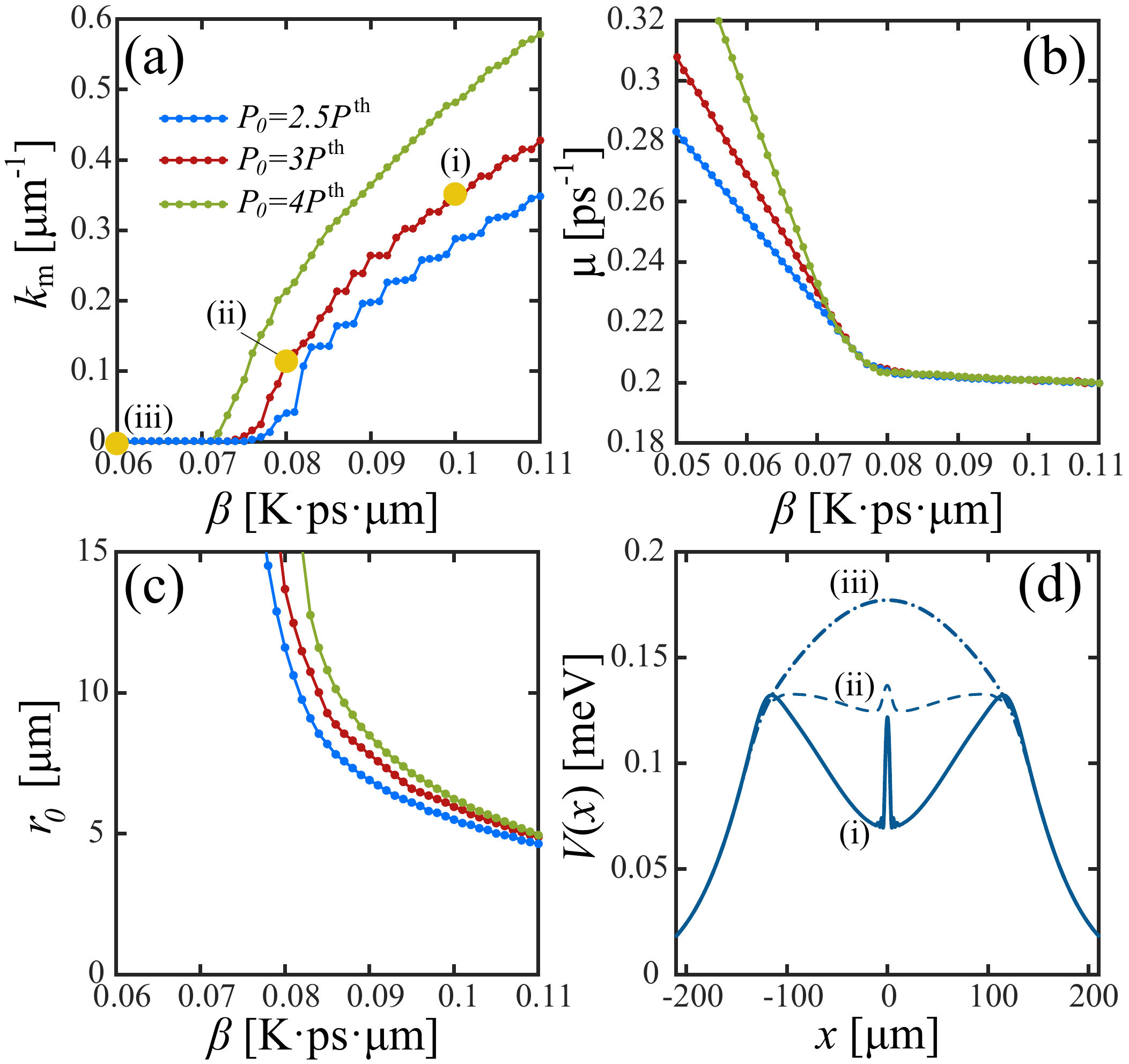}
\caption{Properties of the collective polaron state. Panels (a), (b) and (c) shows the dependencies of the dominant condensate wavevector $k_m$, chemical potential $\mu$ and the polaron localization length on the heating efficiency characterized by $\beta$ parameter for different amplitude of the Gaussian pump \eqref{pump} indicated on (a). (d) The shape of the self-induced nonlinear potential $V(x)$ for three different values of $\beta$ indicated on (a).}
\label{Fig.properties}
\end{figure}

The observed sink-type state supersedes the outflowing solution if the efficiency of heating characterized by   $\beta$ parameter exceeds some critical level, $\beta > \beta_c$.  The spatial spectrum of the steady state clearly illustrates this transition. The spectrum of the sink-type solution is characterized by the two pronounced maxima at $k=\pm k_m$, see Fig.~\ref{Fig.formation}(c), while $k_m=0$ for the outflowing steady state solution. Figure~\ref{Fig.properties} shows the  $k_m(\beta)$ dependence calculated by the multiple numerical solution of the Eq.~(\ref{system}-\ref{pump}) for various values of $\beta$. The value of $k_m$ and thus the magnitude of the polariton flux   increases for $\beta > \beta_c$. Note that the value of $\beta_c$ decreases with the growth of the pump amplitude $P_0$.

The key characteristic of the  polaron state is its localization length which is associated with the density peak width $2r_0$. We estimate this value as a distance between two minima closest to the central peak, see Fig.~\ref{Fig.properties}(a). Since the polaron appears as the result of interference of the incoming polariton fluxes, its width is inversely proportional to the dominant condensate wave vector $k_m$ characterizing the flux. Being delocalized at $\beta \leq \beta_c$ the peak width steeply reduces down to several micrometers as the impact of the heating grows, see Fig.~\ref{Fig.properties}(c).

To reveal the origin of the structure of the observed solution we perform the Madelung transformation of Eq.~(\ref{system}a) substituting $\Psi= \sqrt{\rho(x)} e^{i\phi(x)}  e^{-i\mu t}$ and neglecting energy relaxation term for simplicity:
\begin{subequations}\label{Madelung}
\begin{eqnarray}
\partial_x j  &=& \left[ Rn-\gamma_c\right] \rho,             \\
\hbar \mu  \sqrt{\rho } &=& \left(-\frac{\hbar^2}{2m} \partial_{xx} +  V(x)  + \frac{m}{2}\frac{j^2} {\rho^2}    \right)\sqrt{\rho }  \,
\end{eqnarray}
\end{subequations}
where $V(x) = g_c \rho + g_r n + \alpha(\delta T)$ is an effective nonlinear potential and the flux is $j=\frac{\hbar}{m} {\rho} \partial_x \phi$.

Formation of the bosonic polaron should be attributed to the self-focusing effect originating from the heating of the lattice. If the condensate density is high enough and the heating effect is pronounced, the nonlinear potential $V(x)$ becomes trapping at $\beta > \beta_c$ [see Fig.~\ref{Fig.properties}(d)] and the potential gradient ballistically accelerates the condensate towards the centre. Actually, if the pump intensity  has a smooth profile, the first term in (\ref{Madelung}b) can be omitted. Thus the spatial variation of the nonlinear potential $V(x)$ must be compensated by the flux $j$ growing towards the centre.
Note that the chemical potential $\mu$ of the polaron state is almost independent on $\beta$ for  $\beta>\beta_c$, Fig.~\ref{Fig.properties}(b). It indicates that regardless of the depth of the self-induced trap $V(x)$ the potential energy converts into the kinetic energy of the flowing condensate providing the conservation of the condensate chemical potential \cite{wouters2008}.

Note  that the existence of the converging polariton currents is sustained by the presence of the reservoir, whose density  $n(x)= P(x) \left/ \left(\gamma_r+R|\Psi(x)|^2\right)\right.$    peaks near the pump boundaries, see the solid curve  in the  Fig.~\ref{Fig.formation}(b). According to Eq.~(\ref{Madelung}b) these peaks indicate  the local gain and serve as sources for the polariton fluxes, see Fig.~\ref{Fig.formation}(d). In contrast, for the outflowing state two sinks, i.e. the regions where $\partial_x j < 0$, are located near the pump boundaries.

Even though the self-trapping mechanism favoring formation of the sink-type solution works always provided that the heating is strong enough, the formation of the stable polaron state requires the specific combination of the parameters of the experiment.

\begin{figure}
\includegraphics[width=0.9\linewidth]{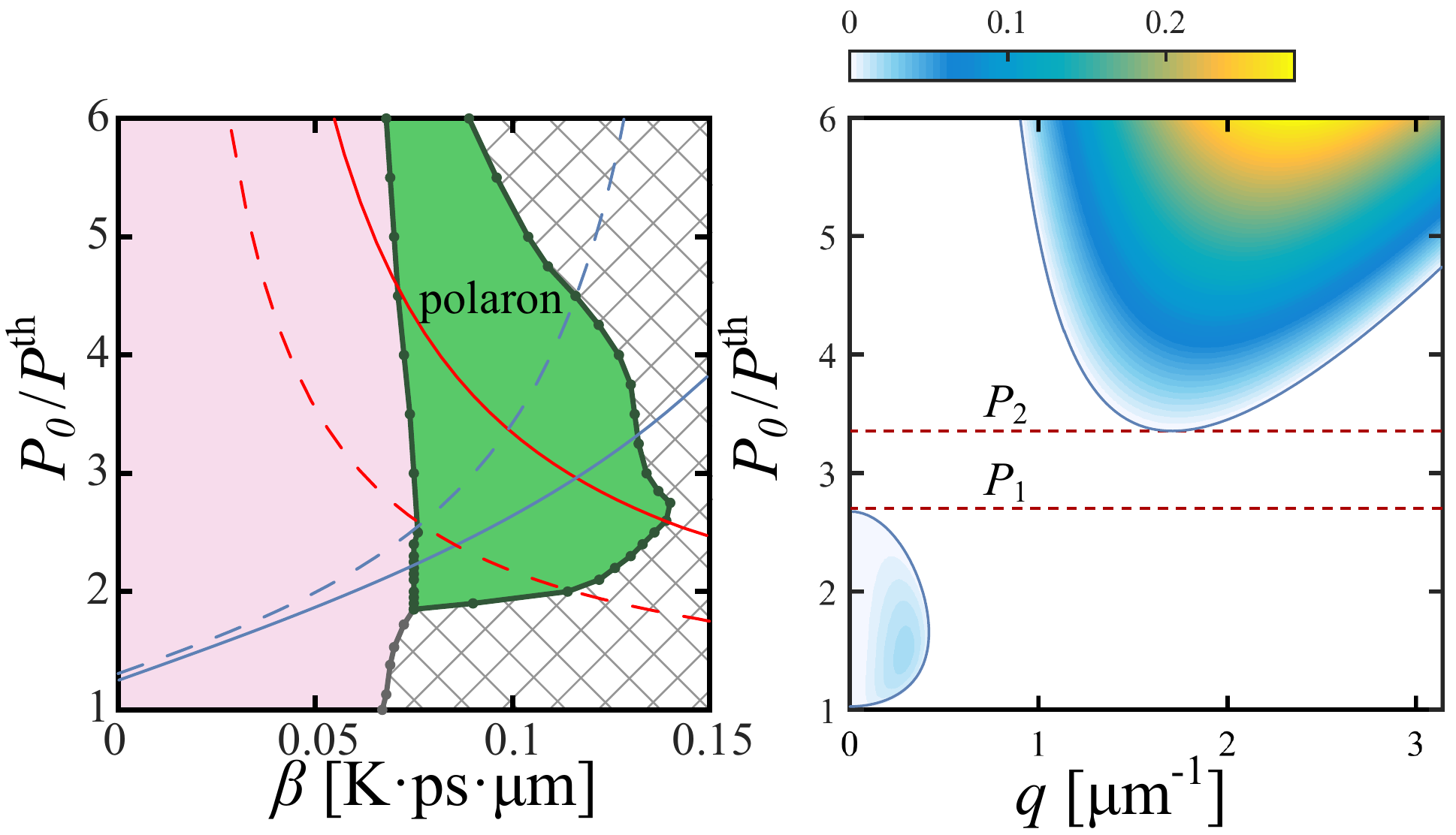}
\caption{(a) The domain of existence of the sink-type polaron state (green shaded region). The bright (pink) region shows the domain where the stable polaron-free solution is established. The hatched region indicates the dynamically unstable states. Red and blue lines indicate the positions of the homogeneous state instability domains $P_2$ and $P_1$, respectively, shown in the panel (b). The solid curves correspond to $R=0.0075$~ps$^{-1}\mu$m while dashed curves correspond to $R=0.0025$~ps$^{-1}\mu$m. (b) The largest linear growth rate (measured in ps$^{-1}$) of the small perturbations of the homogeneous ground state as a function of their momenta $q$ for different pumping intensities $P_0$. }
\label{Fig.domain}
\end{figure}

The domain in the parameter plane $(P_0,\beta)$, where the stable sink-type polaron state is formed, is shaded with green in Fig.~\ref{Fig.domain}(a). The left boundary of this domain corresponds to the critical value $\beta_c$ at which the antitrapping nonlinear potential $V(x)$ changes into the trapping one, see  Fig.~\ref{Fig.properties}(d). Note that the value of $\beta_c$ slightly depends on the pump amplitude. The pink region in the left part of Fig.~\ref{Fig.domain}(a) corresponds to the polaron-free state.

Note that the  polaron state becomes unstable either under the conditions of low pump intensity or at high values of $\beta$. In the latter case the polaron is unstable against weak perturbations and  it is eventually destroyed even if being initially formed at the onset of condensation. The stability properties of the polaron state are  strongly affected by the system parameters.

The position of the right and the bottom boundaries of the polaron stability domain can be estimated analytically in the limit of a large pump spot. In this case the condensate profile slowly varies in space and  its properties are similar to those of a homogeneous infinite condensate  under the homogeneous pump $P(x)=P_0$, for which $|\psi|^2=\left({P_0-P^{\rm{th}}}\right)/{\gamma_c}$ and $n=P_0/\gamma_r$. On the other hand, it is well known that the homogeneous condensate is prone to lose its dynamical stability in a particular range of parameters \cite{smirnov2014,bobrovska2016,gavrilov2018}. So, the loss of stability of the incoming polariton currents should be considered as the main cause preventing formation of the sink-type polaron solution~\cite{saarlos1992,kulczykowski2015}.

The stability properties of the condensate can be analyzed using the Bogolubov-de Gennes \cite{wouters2010} approach which implies perturbation of the steady state and subsequent solution of the eigenvalue problem for the linearized system  \eqref{system}. The details of this calculation are accumulated in the Supplementary materials.

The stability analysis shows that the ground state condensate ($k=0$) is unstable against long wavelength perturbations at the pump power which is above the threshold $P^{\rm th}$ but below some critical value $P_1$. The corresponding linear growth rates of the perturbation with the wave vector $q$ are shown in Fig.~\ref{Fig.domain}(b) in the parameter plane $(P_0,q)$. The heating of the lattice also gives rise to the instability which  evokes excitation of the high-momenta states and occurs for all pump powers above some critical value $P_2$.

The dependencies of the critical pump powers  $P_2$ and $P_1$ on   $\beta$ are shown in Fig.~\ref{Fig.domain}(a) by the  red and the  blue solid curves,  respectively. The ground state is dynamically stable provided that $P_1<P<P_2$, i.e. if $\beta < \beta_{\rm inst} \simeq 0.116$~K$\cdot$ps$\cdot \mu$m for the considered model structure.  The domain of existence of the stable polaron state is slightly wider that is consistent with the previously reported results about  stability of the polariton condensate in the presence of the Gaussian-shaped pump~\cite{bobrovska2014}.

The conditions $\beta>\beta_c$ and $P_2 > P_1$ constitute the criteria of the formation of a sink solution. Note that the value of $\beta_{\rm inst}$ is dependent on the system parameters and may be close to or even smaller than $\beta_c$.  Thus with the reduction of the transition rate $R$  the boundaries of the instability domain shift towards lower value of $\beta$. At the same time the position of $\beta_c$ appears to be almost independent on $R$. For instance, at $R=0.0025$~ps$^{-1}\mu$m (dashed curves in fig.~\ref{Fig.domain}(a)) the stable polaron state does not exist.

In conclusion, we have studied theoretically the self-localization of exciton-polariton condensate due to the heating of the crystal  lattice caused by the relaxation of reservoir excitons into the single quantum state. If the heating efficiency exceeds the critical level, the effective nonlinear trapping potential is formed. Because of its driving-dissipative nature the polariton condensates supports the persistent currents flowing towards the center of the trap where the condensate acquires the soliton-like density peak. The formation dynamics of collective bosonic polarons described here is specific for 1D systems that sustain sink-like solutions of the generalized complex Ginzburg-Landau equations. This finding sheds light on the paradoxical self-trapping effect that seems to contradict the superfluid nature of polariton condensates documented in a number of previous works. We show that there is no controversy here, as the self-localized state is formed due to the interference of two superfluid currents.

Finally, we note that there is an interesting similarity between the collective bosonic polaron described here and bosonic stars formed due to the self-gravity effect in a galactic halo \cite{schunck2003,bekenstein2015}. Based on the formal matching of the generalized Ginzburg-Landau equation considered here and the Gross-Pitaevskii-type equation that governs formation of bosonic stars \cite{navarrete2017}, we cautiously suggest that polariton condensates may serve as a microcrystal laboratory for studies of self-assembly phenomena in cosmology.

This work was supported by the RFBR Grants No. 16-32-60102-mol\_a\_dk and 17-52-10006. The funding from the President of Russian Federation for state support of young Russian scientists, Grant MK-2988.2017.2, and from the Ministry of Education and Science of the Russian Federation, project no. 16.1123.2017/4.6, is acknowledged by I.Yu.Ch.

%

\newpage
\onecolumngrid

\section{Supplementary Material for\\ ``Heat-assisted self-localization of exciton polaritons''}
\subsection{Stability analysis of the homogeneous solution}

Dynamics of the polaron formation is governed by the following coupled equations which account for the heating of the crystal lattice (see Eqs.~(1) and (2) from the main text)
\begin{subequations}\label{system}
\begin{eqnarray}
i\hbar \partial_t \Psi &=& \left[ -\left(\frac{\hbar^2}{2m} - i\Gamma_e \right) \partial_{xx}  + g_c|\Psi^2|  + g_r n   \right. - \alpha_0 \left(2T_0 \beta R n |\Psi|^2  + \beta^2 R^2 n^2 |\Psi|^4\right) +\left. \frac{i\hbar}{2} \left( R n - \gamma_c \right) \vphantom{-\frac{\hbar^2}{2m} \Delta} \right] \Psi,\\
 \partial_t n &=& -\left( \gamma_r + R |\Psi^2| \right)n + P(x).
\end{eqnarray}
\end{subequations}
All the variables are introduced in the main text. If the pump is homogeneous, $P(x)=P_0$, the steady-state solution of these equations is $\Psi(x)=\psi_0 e^{-i \mu t +i k x}$ and $n(x)=n_0$, where
\begin{equation}
|\psi_0|^2=\left({P_0-P^{\rm{th}}}\right)/{\gamma_c}, \ \ \ \ n_0=P_0/\gamma_r,
\end{equation}
$k$ is the momentum characterizing the homogeneous solution, $\mu $ is its chemical potential, $P^{\rm{th}} = \gamma_c \gamma_r \left/ R\right.$ is the condensation threshold. In the ground state, $k=0$ and $\hbar \mu =\frac{\hbar ^2 k^2}{2m} +  \left({g_c - 2T_0 \alpha_0  \beta R n_0}\right)   |\psi_0|^2 +  {g_r}n_0  - \alpha_0 \beta^2 R^2 n_0^2 |\psi_0|^4$.

A stability analysis of the homogeneous solution can be performed by the standard Bogolubov-de Gennes approach \cite{wouters2010}  which implies introduction of the perturbation of the solution in the form
\begin{subequations}\label{stability}
\begin{eqnarray}
 \Psi &=& \left(\psi_0 +\nu e^{i\lambda t/\hbar  + iqx} +  \zeta e^{-i\lambda^\ast t/\hbar  -iqx}\right) e^{-i\mu t + i k x}, \\
 n&=&n_0+ \eta e^{i\lambda t/\hbar +iqx} + \eta^\ast e^{-i\lambda^\ast t/\hbar -iqx} ,
\end{eqnarray}
\end{subequations}
where $q$ is the momentum characterizing weak perturbations $\nu$, $\zeta$ and $\eta$. Linearizing the system (1) from the main text with respect to the perturbations we reduce the stability analysis to the eigenvalue problem $\| {\hat{\cal{L}}}_q-\lambda \hat{E}\|=0$  for the matrix
\begin{equation}{\cal{L}}_q=\left(
           \begin{array}{ccc}
          D^{(+)}_q + N_1 & N_1 & N_2/\psi_0 + \frac{i \hbar}{2}R\psi_0  \\
          -N_1 & - D^{(-)}_q -N_1 & -N_2/\psi_0 + \frac{i \hbar}{2}R\psi_0 \\
          -i\hbar R \psi_0 n_0 & -i \hbar R \psi_0 n_0 & -i\hbar (\gamma_r+R |\psi_0|^2)\\
           \end{array}
         \right)
, \label{stab_matrix}
\end{equation}
where $\hat{E}$ is the identity matrix, $N_1= g_c|\psi_0|^2 - \alpha_1 n_0 |\psi_0|^2 - 2\alpha_2 n_0^2 |\psi_0|^4$, $N_2 = g_r |\psi_0|^2 -\alpha_1 |\psi_0|^4 -2\alpha_2 n_0 |\psi_0|^6$ with $\alpha_1 = 2 \alpha_0 T_0 \beta R$ and $\alpha_2 = \alpha_0 \beta^2 R^2$. $D^{\pm}_q = \left(\frac{\hbar^2}{2m} \mp i\Gamma_e \right)\left( q^2 \pm 2kq \right)$ are the linear dispersions of the perturbations with the momenta $k+q$ and $k-q$, respectively. The solution is stable if the imaginary part $\lambda_i$  of the eigenvalue $\lambda=\lambda_r+i\lambda_i  $ is positive for any $q$.

\begin{figure}
\includegraphics[width=1\linewidth]{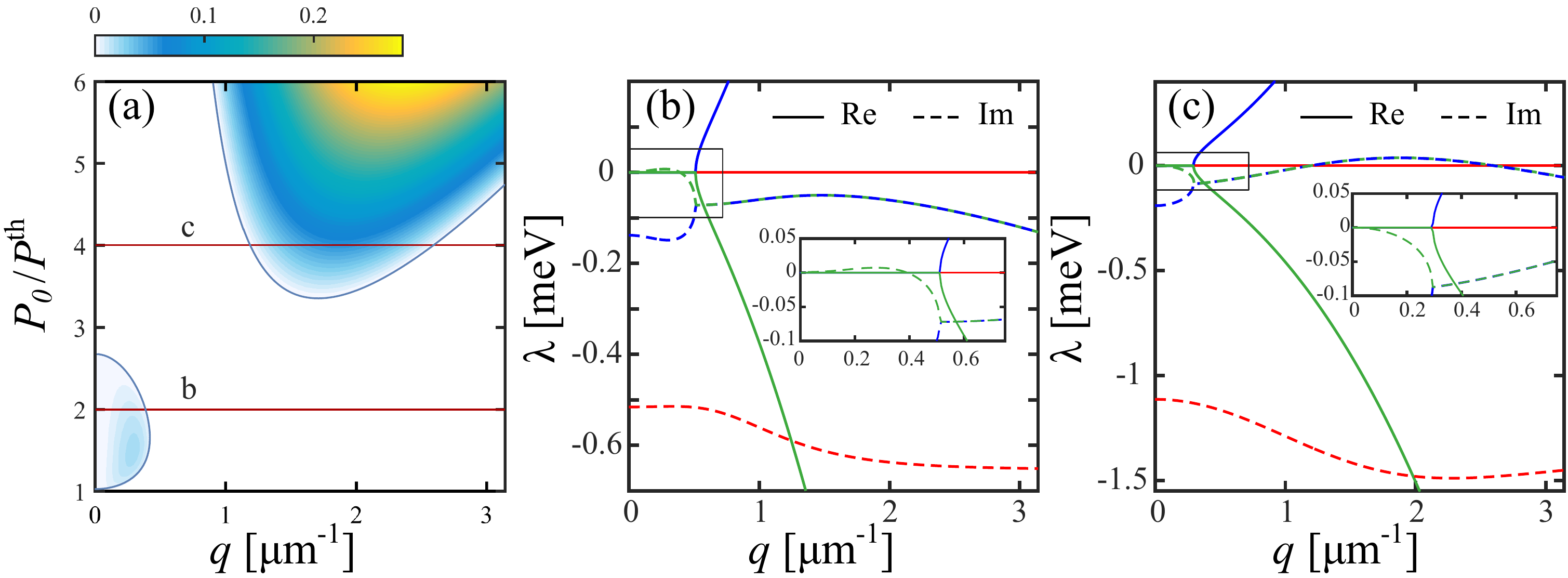}
\caption{(a) The largest linear growth rate (measured in ps$^{-1}$) of the small perturbations of the homogeneous ground state as a function of their momenta $q$ for different pumping intensities $P_0$. (b) and (c) The Lyapunov spectra of the condensate ground state formed by the homogeneous pump with amplitudes (b) $P_0=2 P^{{\rm th}}$ and (c) $P_0=4 P^{{\rm th}}$. Solid curves show to the real part of the perturbation frequency $\lambda$ while dashed curves show the imaginary part. The insets  show  magnified regions  in the frames.}
\label{SupplFig.1}
\end{figure}

The dispersion of the real and imaginary parts of the Lyapunov exponent $\lambda(q)$ in different regimes are shown in Fig.~\ref{SupplFig.1}. The instability associated with the Goldstone mode whose dispersion tends to zero at $k=0$ appears under the moderate pump powers which are close to the threshold, see Fig.\ref{SupplFig.1}.
The upper limit $P_1$  of the pump powers supporting the instability (see Fig.~\ref{SupplFig.1}(a)) can be obtained from the characteristic cubic equation corresponding to the eigenvalue problem for the matrix \eqref{stab_matrix}.
Since the stability of the considered solution is governed by the change of the sign of the imaginary part $\lambda_i$ of the perturbation frequency, the terms  containing $\lambda_i$ can be omitted. In this case the characteristic equation reduces to the following conditions:
\begin{subequations}\label{conditions}
\begin{eqnarray}
\lambda_r^3 + B \lambda_r=0,\\
A\lambda_r^2+D=0,
\end{eqnarray}
\end{subequations}
where
$A=\hbar P_0 / n_0 - 2\Gamma_e q^2$,
$B=\left(\frac{\hbar^2}{m} N_1 + 2\hbar \Gamma_e P_0/ n_0 \right) q^2 + \left( \frac{\hbar^4}{4m^2}+\Gamma_e^2 \right)q^4 + \hbar^2 R^2 |\psi_0|^2 n_0$, and
$D=\left(\frac{\hbar^3}{m} \left( N_1 P_0/n_0- R n_0 N_2\right) + \hbar^2 \Gamma_e R^2 |\psi_0|^2 n_0 \right)   q^2 + \hbar P_0/n_0\left( \frac{\hbar^4}{4m^2}+\Gamma_e^2 \right)q^4 $ .
The Eqs.~\eqref{conditions} have a solution $\lambda_r=0$ and $D=0$. The latter condition determines the boundary of the lower instability domain in the $(P,q)$ plane in Fig.~\ref{SupplFig.1}(a). The value of $P_1$ is determined from the single root of the $P(q)$ plane dependence corresponding to $q=0$ \cite{smirnov2014}. In this case from the condition $D=0$ one can obtain:
\begin{equation}\label{P1}
P_1= \gamma_c \frac{g_r \gamma_c  +  P^{\rm{th}} \left(\alpha_1 -2\alpha_2  P^{\rm{th}} /R\right) - \Gamma_e m \gamma_c/\hbar}{Rg_c - 2\alpha_2  P^{\rm{th}} \gamma_c/R}.
\end{equation}
Under $\alpha_0=0$ and $\Gamma_e=0$ this expression reduces to the well-known criterion of the modulational instability of the nonresonantly pumped polariton condensate ground state: $ P_0/P^{\rm{th}} <  {\gamma_c g_r}/{\gamma_r g_c} $, \cite{smirnov2014,liew2015,bobrovska2016}.

Note that the instability associated with the Goldstone mode involves excitation of the long-wave-length perturbations. Thus in the system with the finite-size pump the instability appears only if the pump profile is smooth and the pump width is comparable with the perturbation wavelength.

Accounting for the heat released during condensation gives rise to the instability of a new type which evokes excitation of the high-momenta states (see Fig.~\ref{SupplFig.1}(c)) and occurs for all continuous pump powers above $P_2$. The boundary of the domain associated with this instability can be also determined from the system \eqref{conditions}. The second solution of Eqs.~\eqref{conditions} corresponds to $\lambda_r^2=-B$ and requires negative values of $B$. The latter becomes possible provided that $N_1<0$, i.e. if the heat-assisted energy red shift $\alpha_1 n_0 |\psi_0|^2 + 2\alpha_2 n_0^2 |\psi_0|^4$ dominates the blue shift from the polariton-polariton repulsion $g_c|\psi_0|^2$. The boundary of the instability domain in the $(P,q)$ plane obeys  the following parametric equation which follows from Eq.~(\ref{conditions}b):
\begin{eqnarray}\label{P2}
\frac{2\Gamma_e}{\hbar} \left( \frac{\hbar^2}{4m^2} +\frac{\Gamma_e^2}{\hbar^2} \right)q^6 + \left( \frac{2\Gamma_e N_1}{\hbar m} + 4\frac{\Gamma_e^2 RP_0}{\hbar^2 \gamma_c} \right)q^4 +\left( 2\frac{R^2P_0^2\Gamma_e}{\hbar \gamma_c^2} + \frac{\gamma_c N_2}{m} + \frac{\Gamma_e}{\hbar}P_0(P_0-P^{{\rm th}}) \right)q^2 + \frac{R^2}{\gamma_c}P_0\left(P_0-P^{\rm{th}} \right) = 0.
\end{eqnarray}
The value of $P_2$ corresponds to the minimum of   $P(q)$ dependence obeying  $P_2>P^{\rm{th}}$. It indicates the maximum pump strength supporting a stable ground state solution provided that $P_2>P_1$.

\end{document}